\documentclass[10pt,a4paper,twocolumn,showpacs,amsmath,pra,amssymb,english]{revtex4}

\usepackage[T1]{fontenc}
\usepackage[latin9]{inputenc}
\usepackage{float}
\usepackage{graphicx}
\usepackage{amssymb}
\usepackage{esint}
\usepackage{times}

\makeatletter

\usepackage{babel}
\makeatother

\begin{document}

\title{Analysis of the Holzmann-Chevallier-Krauth theory for the trapped quasi-two-dimensional
Bose gas}

\author{R.~N.~Bisset, D.~Baillie and P.~B.~Blakie}
\affiliation{Jack Dodd Centre for Quantum Technology, Department of Physics, University of Otago, New Zealand}


\date{\today}
\begin{abstract}
We provide an in depth analysis of the theory proposed by Holzmann, Chevallier and Krauth (HCK) [Europhys.~Lett.,  {\bf 82}, 30001 (2008)] for predicting the temperature at which the Berezinskii-Kosterlitz-Thouless (BKT) transition to a superfluid state occurs in the harmonically trapped quasi-two-dimensional (2D) Bose gas. Their theory is based on a meanfield model of the system density and we show that the HCK predictions change appreciably when an improved meanfield theory and  identification of the transition point is used. In this analysis we develop a consistent theory that provides a lower bound for the BKT transition temperature in the trapped quasi-2D Bose gas.
\end{abstract}

\maketitle
\section{Introduction}
Thermal fluctuations destroy long-range coherence
in the  two dimensional (2D) Bose gas  \cite{Hohenberg1967,Posazhennikova2006}.
However,  a Berezinskii-Kosterlitz-Thouless  (BKT)  transition \cite{Berezinskii1971,Kosterlitz1973} to a quasi-coherent superfluid state is predicted to occur, and was observed first in  liquid helium thin
films in 1978 \cite{Bishop1978}.
More recently evidence for this BKT transition was reported for a
dilute Bose gas by the ENS and NIST groups \cite{Stock2005,Hadzibabic2006,Kruger2007,Clade2008} in harmonically trapped quasi-2D systems.

 These experiments raise two important issues that need to be dealt with:
(i) The trapping potentials are not purely two-dimensional for the temperature regimes considered, with some thermal excitation in the tight direction.
  (ii) The weak harmonic confinement in the 2D plane of the system introduces finite size effects, and competition between interaction and potential energy of the system.
These issues have made the low temperature
phase diagram of this system the subject of much debate \cite{Popov1983,Petrov2000,Prokofev2001,Prokofev2002,Andersen2002,Gies2004,Trombettoni2005,Simula2005}, with reliable predictions only coming recently from classical field and quantum Monte Carlo
methods \cite{Simula2006,Holzmann2007b, Simula2008a,Bisset2008}.

It is of course desirable to have a simple meanfield description of the quasi-2D system  \cite{Hadzibabic2008a,Holzmann2008a}. However, meanfield theories are of limited applicability in the critical region, where density fluctuations are strong, and it is well known that the 2D critical region is large.
However, recently Holzmann, Chevallier and Krauth (HCK) \cite{Holzmann2008a} made a novel proposal to use a high temperature Hartree-Fock meanfield theory to extrapolate into the lower temperature critical regime. They then used this theory to estimate the transition temperature ($T_{BKT}$) as that where the peak phase space density of the system satisfies the critical value (see Eq. (\ref{ncrit})) known for the uniform pure-2D Bose gas \cite{Prokofev2001,Prokofev2002}.

In this paper our principle focus is to analyze two aspects of the HCK theory: (i) The simplification of the interaction term to an averaged value that is insensitive of the particular axial mode the atoms occupy. (ii) The use of the total areal density to identify the BKT transition.
In our analysis we use a more complete high temperature meanfield theory that avoids the interaction simplification used by HCK.
We also show that the correct generalization of the pure-2D condition for the  BKT transition to the quasi-2D system involves the areal density of the ground axial mode of the system.
Through numerical calculations we show that our improved treatment of these two aspects leads to significant differences in our theoretical predictions from those of HCK.
We also discuss the main limitation of meanfield theory extrapolation into the critical regime, which indicates that our improvements on the HCK theory will provide a lower bound on the temperature for $T_{BKT}$ in the trapped quasi-2D system.

\section{Formalism}
We begin with the effective low-energy Hamiltonian for ultra-cold bosonic atoms
\begin{eqnarray}
\hat{H} &=& \int{d^3\mathbf{x}\,\hat{\Psi}^\dag \left(-\frac{\hbar^2}{2m} \nabla^2+\frac{1}{2}m\sum_j\omega_j^2x_j^2\right) \hat{\Psi}} \nonumber\\&&+ \frac{2\pi a \hbar^2}{m}\int{d^3\mathbf{x}\,  \hat{\Psi}^\dag\hat{\Psi}^\dag  \hat{\Psi} \hat{\Psi}},\label{Hfull}
\end{eqnarray}
where $\hat{\Psi}({\mathbf{x}})$ is the quantum Bose field operator that annihilates a particle at position ${\mathbf{x}}$, $a$ is the s-wave scattering length  and $m$ is the atomic mass.
The quasi-2D system we consider here is realized when the trapping potential is sufficiently tight in one direction (which we take to be $z$)  that $\hbar\omega_{x,y}\ll k_BT\sim\hbar\omega_z$.

Our interest is in the thermal properties of the quasi-2D system when there is no condensate present, a regime for which Hartree-Fock theory is appropriate.
If interactions are small compared to $\hbar\omega_z$ then the Hartree-Fock modes for the Hamiltonian (\ref{Hfull}) take the separable form $\psi_{k\sigma}(x,y,z)=f_{k\sigma}(x,y)\xi_k(z)$, where the axial modes $\xi_k(z)$ are bare harmonic oscillator states.
In the quasi-2D regime the $xy$-plane, for which we introduce the notation $\mathbf{r}=(x,y)$, can be treated semiclassically, eliminating the need to diagonalize for the modes $f_{k\sigma}(x,y)$. However, the axial modes must be treated quantum mechanically, and the Hartree-Fock expression for the areal density of the system in the $j$-th axial mode is
\begin{equation}\label{nj}
n_j(\mathbf{r}) = \frac{1}{(2\pi)^2}\int d^2\mathbf{k}_r\, \frac{1}{\exp{\left\{\frac{\epsilon_j(\mathbf{r},\mathbf{k}_r)-\mu}{k_BT}\right\}} - 1},
\end{equation}
where the Hartree-Fock energies are
\begin{eqnarray}
\epsilon_j(\mathbf{r},\mathbf{k}_r) &=& \frac{\hbar^2k_r^2}{2m} + \frac{m}{2}(\omega_x^2x^2+\omega_y^2y^2)+ j\hbar\omega_z\label{ej}\\
 && + 2\sum_{k=0}^\infty g_{kj}n_k(\mathbf{r}),\nonumber
\end{eqnarray}
$\mu$ is the chemical potential, and
\begin{equation}
g_{ki} =\frac{4\pi a\hbar^2}{m}\int dz \,|\xi_k(z)|^2|\xi_i(z)|^2,
\end{equation} describes the interactions between atoms in the $k$ and $i$ axial modes.
Performing the momentum integration in Eq.~(\ref{nj}) and adding up the axial mode densities gives the total areal density
\begin{equation}\label{localden}
n(\mathbf{r}) = -\frac{1}{\lambda^2}\sum_{j=0}^\infty\ln \left[1-\exp\ \left(\{\mu-V_j(\mathbf{r})\}/k_BT\right)  \right],
\end{equation}
where
\begin{equation}
V_j(\mathbf{r})=\frac{m}{2}(\omega_xx^2+\omega_yy^2) +j\hbar\omega_z+2\sum^\infty_{k=0}g_{kj}n_k(\mathbf{r}),
\end{equation}
is the effective potential for atoms in the $j$-th axial mode,
and $\lambda = h/\sqrt{2\pi mk_BT}$ is the thermal de Broglie wavelength.
We solve Eqs.~(\ref{nj}) and (\ref{ej}) self-consistently, i.e.~by iterating until the solutions converge.
\subsection{Comparison to theory of Holzmann et al.}
A central concern of our work is to compare our meanfield theory, as outlined above, to the meanfield theory used by HCK \cite{Holzmann2008a}. The HCK theory is a simplification of our meanfield scheme presented above made by taking the interactions to be axial mode independent,
 i.e.,  changing the meanfield interaction term to
\begin{equation}\label{HCKinter}
2\sum_{k=0}^\infty g_{kj}n_k(\mathbf{r})\rightarrow
2g_{\rm{H}}n(\mathbf{r}).
\end{equation}
This approximation has no rigorous justification, but allows a closed-form expression for total density.
The \emph{average} interaction strength used in the HCK model is given by
\begin{equation}
g_{\rm{H}}=\frac{4\pi a \hbar^2}{m}\int dz [ \rho (z)]^2,
\end{equation}
where $\rho(z)$ is the density of a single atom in a harmonic oscillator of frequency $\omega_z$ at temperature $T$.

\subsection{Results: 2D phase space density}\label{sec2DPSden}
\begin{figure}
\includegraphics[width=2.8in]{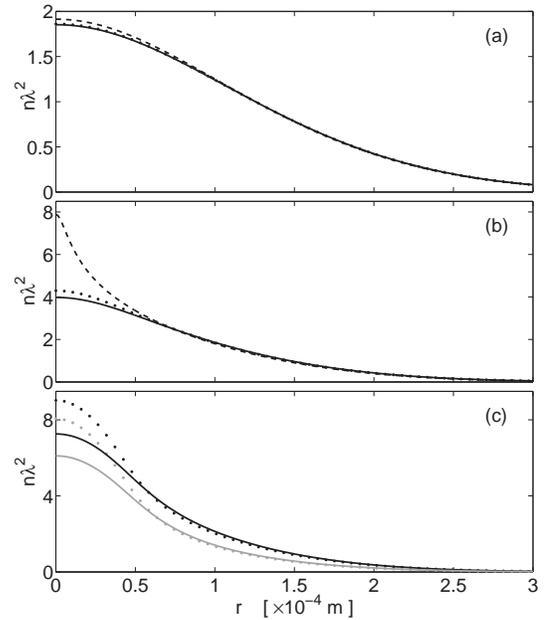}
\caption{\label{fig:denprof} Comparison of 2D phase space densities for systems at  (a) $T = 221$ nK, (b)  $T=172$ nK, and (c) $T=150$ nK. Our meanfield model (solid), HCK model (dotted) and ideal gas (dashed). Ground axial mode areal densities are shown in grey in (c). Calculation parameters are $N = 10^4$ ${^{87}}$Rb atoms with $\omega_{x,y} =2\pi\times59$ Hz and $\omega_z = 2\pi\times2530$ Hz. We note that the parameters in (a) are the same as those in Fig.~2 of \cite{Holzmann2008a}.}
\end{figure}

In Fig.~\ref{fig:denprof} we compare the two meanfield models and the ideal gas model (i.e.~Eqs.~(\ref{nj}) and (\ref{ej}) with $a=0$) for three different temperatures.
The density profiles shown in Fig.~\ref{fig:denprof}(a) are at the pure-2D condensation  temperature\cite{Bagnato1991}, i.e.~$T^{2D}_{BEC} \equiv \sqrt{6N}\hbar\sqrt{\omega_x\omega_y}/\pi k_B$ \footnote{We note that interaction effects suppress condensation and we only indicate the Bose-Einstein condensation temperature to indicate why the ideal gas shows saturation.}. In this regime both meanfield models make similar predictions, and have lower density than the ideal gas at the trap center due to the effects of repulsive interactions.

The results in Fig.~\ref{fig:denprof}(b) are at a colder temperature where the ideal gas is almost saturated. (i.e.~approximately at the quasi-2D condensation  temperature, $T^{q2D}_{BEC}<T^{2D}_{BEC}$, where the central density begins to diverge). Interaction effects play a more significant role here, and prevent the density from spiking in both meanfield theories. In this regime, and at lower temperatures [Fig.~\ref{fig:denprof}(c)], the differences between the meanfield theories become clearly apparent.
\begin{table}
\begin{tabular}{l | c |c c c| c c}
\hline\hline
 $T [nK]$ & $g_{\rm{H}}\times\frac{m}{\hbar^2}$  & $N_0$ $$ & $N_0$ $^{\rm{(HCK)}}$ & $N_0$ $^{\rm{(Boltz)}}$ & $(n\lambda^2)_{\rm{peak}}$  & $(n_0\lambda^2)_{\rm{peak}}$ \\
\hline
132 & $0.0851$ &  $7666$  & $7888$ & $6024$ &  $10.4$ & $9.38$\\
150 & $0.0805$ &  $6815$  & $7010$ & $5559$ & $7.24$ & $6.08$\\
 172 & $0.0756$ & $5834$ & $5928$ &  $5073$ & $3.96$ & $2.76$ \\
 221 & $0.0673$ &  $4560$  & $4531$ & $4236$ & $1.85$ & $0.91$\\
 270 & $0.0612$ &  $3774$  & $3788$ & $3630$ & $1.17$ & $0.46$\\
 \hline\hline
\end{tabular}

\caption{\label{paramtable}Comparison of parameters and theoretical predictions for the quasi-2D system considered in Fig.~\ref{fig:denprof}.
$N=10^4$ $^{87}$Rb atoms. Interaction parameters, values of the ground axial mode occupation $N_0=\int d^2\mathbf{r}\,n_0(\mathbf{r})$ for our theory ($N_0$),  HCK theory ($N_0^{\rm{(HCK)}}$), and the Boltzmann case ($N_0^{\rm{(Boltz)}}$). The peak (areal) total phase space and ground axial mode phase space densities of our theory are shown. In our theory the first few interaction parameters are: $\{g_{00},g_{01},g_{11}\}=(\hbar^2/m)\{0.1299,0.0650,0.0974\}$.}
\end{table}
This discrepancy arises from how interactions are treated in the  theories  [see Eq.~(\ref{HCKinter})] in two ways:\\
 {\bf (i) Averaged interaction parameter:} The  interaction parameter, $g_{\rm{H}}$, in the HCK theory assumes that the various modes follow a Boltzmann distribution. As the system becomes degenerate, $n_0\lambda^2\sim1$, this approximation is inaccurate as it fails to account for quantum statistical effects that increase the ground mode occupation.  A comparison of both meanfield theories  and the Boltzmann prediction of the ground band occupation are given in Table \ref{paramtable}, and reveals the increasing difference between the meanfield and Boltzmann results as the phase space density increases.\\
{\bf (ii) Mode independence of interaction:} The marked difference between the ideal and meanfield solutions [e.g.~see Fig.~\ref{fig:denprof}(b)] arises from interaction effects.  These effects are dominated by the atoms in the ground axial mode.
Because $g_{\rm{H}}<g_{00}$, our ground mode atoms are more strongly interacting than those in the HCK theory, thus our theory predicts  $n_0(\mathbf{r})$ to be lower and broader  [see Fig.~\ref{fig:denprof}(c)],  with
less atoms in the ground axial mode [see Table \ref{paramtable}].

Finally, we comment on validity aspects of our Hartree-Fock approach (also see Sec. \ref{MFvalid2}).
First, the separability of our Hartree-Fock description into planar ($f_{k\sigma}(\mathbf{r})$) and axial (harmonic oscillator, $\xi_k(z)$) modes requires that interaction effects are small compared to the axial energy scale, i.e. $n_0(\mathbf{0})g_{00}/\hbar\omega_z\ll1$.
 For the results presented in Fig.~\ref{fig:denprof}, the value of this ratio varies from $0.034$ (Fig.~\ref{fig:denprof}(a)) to $0.16$ (Fig.~\ref{fig:denprof}(c)).

\section{ BKT transition in the trapped system}
\subsection{Monte Carlo analysis of the BKT transition}
The BKT transition occurs when the superfluid density, $n_{\rm{SF}}$, satisfies $n_{\rm{SF}}\lambda^2=4$ \cite{Kosterlitz1973}.
Monte Carlo calculations by Prokof'ev et al., \cite{Prokofev2001} have characterized the uniform 2D Bose gas, showing that this condition, in terms of the total density, is \begin{equation}\label{ncrit}
n^{\rm{crit}}_{2D} = \frac{1}{\lambda^2}\ln\left(\frac{\hbar^2C}{mg}\right),
\end{equation}
where $C=380\pm3$, ${g}$ is the 2D interaction strength \cite{Prokofev2001, Prokofev2002}, and the subscript $2D$ emphasizes that this result is for the pure-2D Bose gas.
An important point made in Ref.~\cite{Prokofev2001} is that the long wavelength behavior of all 2D weakly interacting  $|\psi|^4$ models is universal at the transition point. However, differences between models emerge in high energy modes, which contribute to the total density at the critical point, but do not affect the strongly fluctuating critical region. Hence, we can add or subtract meanfield contributions for the high energy modes which modifies the total density and the parameter $C$ used in Eq.~(\ref{ncrit}). The value $C\approx380$ only applies to the continuous Bose gas.

The uniform criterion (\ref{ncrit}) can be applied to the quasi-2D trapped gas using a local density approximation. Within such an approximation the transition will be spatially dependent, and will occur first at the centre of the trap where the density is highest. Some care has to be taken in extending the pure-2D theory to the quasi-2D case, since the additional axial modes correspond to a different meanfield theory.
The correct quasi-2D extension of result (\ref{ncrit}) should  be applied to the subsystem consisting of atoms occupying the ground axial mode, with relevant interaction parameter $g_{00}$, i.e.
\begin{equation}\label{ncrit2}
n^{\rm{crit}}_{0} = \frac{1}{\lambda^2}\ln\left(\frac{\hbar^2C}{mg_{00}}\right),
\end{equation}
where  $C\approx380$.
\subsubsection{Meanfield validity near the transition}\label{MFvalid2}
We note that when condition (\ref{ncrit2}) is satisfied the meanfield theory must be inapplicable as a complete description. However, the meanfield theory itself predicts no transition to BKT (or BEC) phases and varies smoothly through this temperature range. Thus the basis of the HCK approach, the use of meanfield  theory to estimate the system phase space density, would seem to be a reasonable starting point for estimating $T_{BKT}$ in the trapped system.  The main deficiency of the meanfield approach is that it assumes Gaussian fluctuations, whereas near the transition these are suppressed via the formation of a quasi-condensate \cite{Bisset2008,Prokofev2001}.   In such regimes the Hartree-Fock treatment overestimates the energetic cost due to interactions, and predicts the system to spread out more and hence have a lower central density. We hence conclude that near the BKT transition our meanfield density will be less than the actual system density.
As discussed in Sec. \ref{sec2DPSden}, the simplifying assumption of the HCK theory leads to a lower self interaction for the ground mode which increases the density, though for reasons unrelated to quasi-condensation.
Thus the HCK theory cannot be assured to provide a lower bound for the system density.


\subsection{Using meanfield theory to predict the BKT transition}\label{secBKTMFpredictions}
While both Hartree-Fock meanfield theories discussed fail to predict any transition in the quasi-2D Bose gas, their predictions for the density can be extrapolated \footnote{We use the term extrapolate to emphasize that these theories are not valid at the transition as quasi-condensation will significantly effect the density at this point} to estimate when the quasi-2D condition for the BKT transition is fulfilled (\ref{ncrit2}).
 This approach was formulated in \cite{Holzmann2008a}, but using the total areal density, $n$, and the averaged interaction parameter, $g_{\rm{H}}$,  in  expression (\ref{ncrit}). These choices are in contrast to the correct quasi-2D extension of the universal result given in Eq.~(\ref{ncrit2}).

\subsection{Comparison of meanfield predictions for the BKT Transition}
\begin{figure}
\includegraphics[width=2.8in]{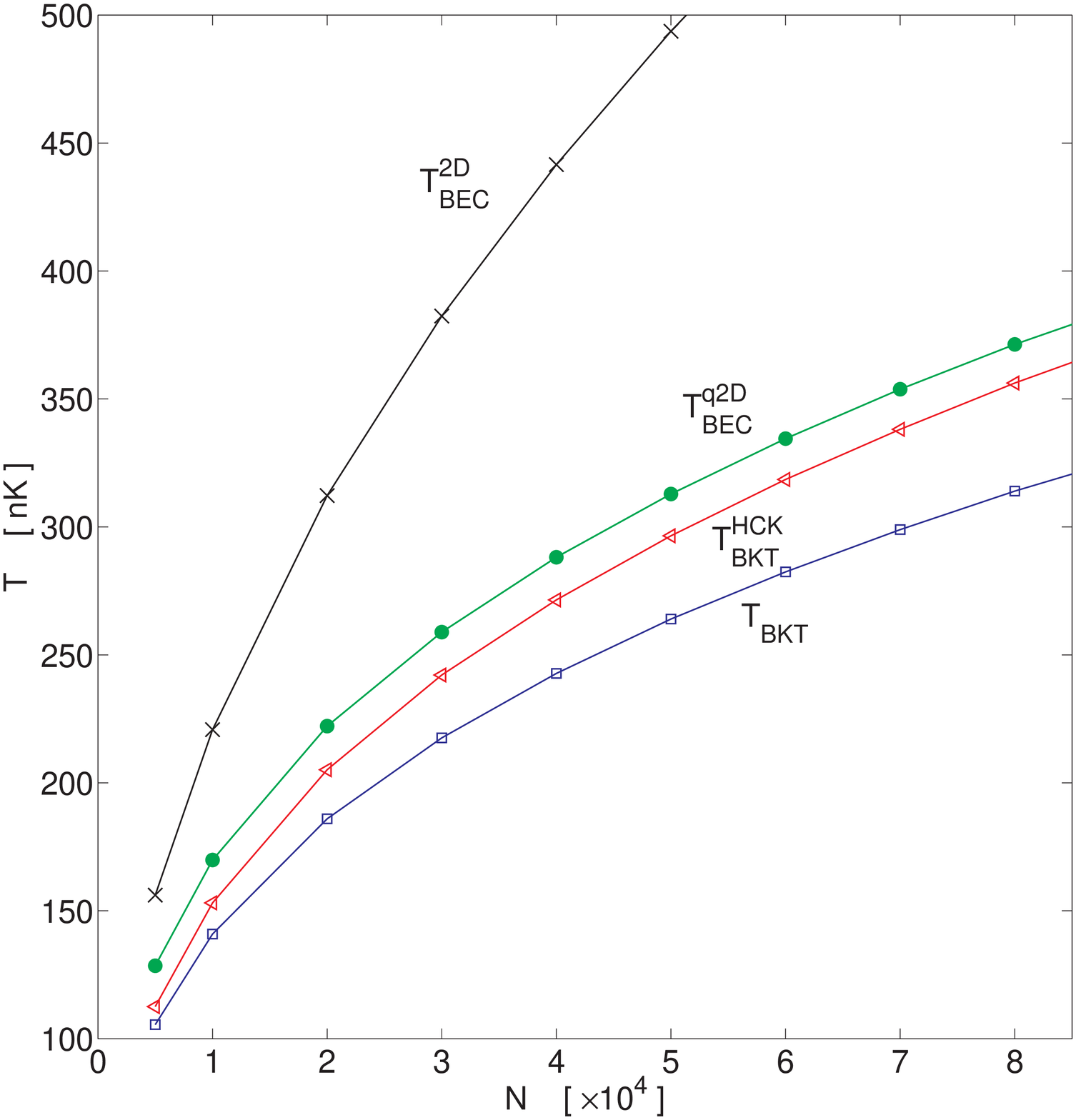}
\caption{\label{fig:semiclasca} Predictions for the BKT transition temperature as a function of $N$. Our theory (squares), HCK theory (triangles), quasi-2D BEC (points) and pure 2D BEC temperature (crosses). All other parameters as in Fig.~\ref{fig:denprof}.}
\end{figure}
We now explore the difference in the  meanfield-extrapolated predictions for the BKT transition temperature ($T_{BKT}$) between (i) our meanfield theory using Eq.~(\ref{ncrit2}) and (ii) the HCK theory using
$n^{\rm{crit}} = {\lambda^{-2}}\ln\left({\hbar^2C}/{mg_{\rm{H}}}\right)$.
These predictions are shown in Fig.~\ref{fig:semiclasca} for systems of various total particle number.
Over the entire range of atom numbers considered our $T_{BKT}$ estimates are significantly lower than those of the HCK theory, indeed, for the system with $8\times10^4$ atoms, our prediction is approximately $50$ nK lower. However, we note that for the largest atom numbers considered,  the ratio of $k_BT_{BKT}/\hbar\omega_z$ is sufficiently large that the system is crossing over to the 3D regime.

A basic comparison of the two Hartree-Fock theories has already been discussed in Sec.~\ref{secBKTMFpredictions}, however two main factors are responsible for the appreciable differences in their extrapolated-predictions of $T_{BKT}$:


{\bf (i) Smaller interaction parameter:} The interaction parameter used by HCK to determine the critical density  (i.e.~$g_{\rm{H}}$) is smaller  than ours ($g_{00}$) [see Table \ref{paramtable}]. Noting that the interaction parameter only effects the critical density logarithmically (\ref{ncrit2})), this difference leads to a negative shift in the HCK prediction for the critical temperature.

{\bf (ii) Use of meanfield density:} Using the total areal density to judge when the system is critical causes the HCK theory to predict a higher critical temperature than our approach which instead uses $n_0(\mathbf{r})$.  Furthermore the HCK meanfield theory predicts higher densities through the use of the averaged interaction parameter, as discussed earlier.  A comparison of the various areal densities of these two theories in the critical regime for $N=10^4$ atoms is given in Fig.~\ref{fig:denprof}(c).

Effect (ii) dominates for the results in Fig.~\ref{fig:semiclasca}, increasingly so for larger numbers of atoms.

\section{Conclusions}

In this paper we have outlined a systematic Hartree-Fock meanfield theory for the trapped quasi-2D Bose gas, which should provide a good description above the BKT critical region.
A central concern of this paper has been to compare our theory against that of HCK \cite{Holzmann2008a}, which involves an unjustified simplification of the meanfield interaction term.
We show that the density profiles predicted by these theories disagree as the temperature decreases.

We have also considered how to extend the BKT critical density condition
to the trapped quasi-2D Bose gas.  Using this extension we are able to extrapolate our meanfield theory to predict the critical temperature.
We show that the HCK extrapolation, which uses a different local density condition to identify the BKT critical point, predicts a significantly higher value for $T_{BKT}$.
%

Finally, we would like to emphasize that our treatment for extrapolating meanfield theory to predict the BKT transition will likely under-estimate $T_{BKT}$. This arises because the assumed Gaussian fluctuations of the Hartree-Fock theory reduce the system density compared to that of the actual system, which forms a quasi-condensate as a precursor to the BKT transition. Thus our theory will provide a lower bound on where the system will obtain the critical density expected for the BKT transition to occur.

\section*{Acknowledgments}
The authors acknowledge many useful discussions with T.~P.~Simula and D.~A.~W.~Hutchinson. We thank M.~Holzmann for his feedback on this manuscript.
This work was financially supported by the University of Otago and the New Zealand Foundation for Research Science and Technology under the contract NERF-UOOX0703: Quantum Technologies.

\bibliography{reference}





\end{document}